\documentclass[a4paper,aps,twocolumn]{revtex4}
\RequirePackage[english]{babel}
\RequirePackage[latin1]{inputenc}
\RequirePackage[T1]{fontenc}
\RequirePackage{mathrsfs}
\RequirePackage{amsmath}
\RequirePackage{amssymb}
\RequirePackage{amsbsy}
\RequirePackage{bm}
\begin{document}
\title{\bf{Least-Order Torsion-Gravity for Chiral-Spinor Fields, induced\\ 
Self-Interacting Potentials and Parity Conservation}}
\author{Luca Fabbri}
\affiliation{INFN \& Dipartimento di Fisica, Universit{\`a} di Bologna,\\
Via Irnerio 46, 40126 Bologna and\\
DIPTEM, Universit\`{a} di Genova,\\
Piazzale Kennedy Pad. D, 16129 Genova, ITALY}
\date{\today}
\begin{abstract}
We will consider the most general least-order derivative action for the torsional completion of gravitational backgrounds filled with left-handed and right-handed semi-spinorial fields, accounting for all parity-even as well as parity-odd contributions; we will proceed by performing the customary analysis, decomposing torsion and substituting it in terms of the semi-spinorial density currents, in order to obtain the effective action with the torsionally-induced self-interacting potentials among the chiral fermionic fields: we shall see that the resulting effective non-linear potentials will turn eventually out to be parity conserving after all.
\end{abstract}
\maketitle
\section*{Introduction}
In the developments of the foundations of theoretical physics, the construction of a specific model is based on a simple prescription, that is, after building the geometric background in which to host matter fields, one has to require a coupling between the underlying space-time and its matter content by defining the field equations, or in modern terms, the dynamical action; of course, even if we restrict ourselves to $(1\!+\!3)$-dimensional space-times and additionally to the $\frac{1}{2}$-spin spinor fields, there still exists an infinite number of such dynamical actions, unless we fix some additional constraint, and in this paper, our assumption will be to restrict ourselves to least-order derivative dynamical actions solely: this constraint makes the dynamical actions rather simple, and so one may reasonably look for the most general of them all.

The problem of finding the most general least-order derivative dynamical action for the torsional completion of gravity of an underlying space-time filled with Dirac spinor fields has been addressed in the literature, and the most recent results were reported in reference \cite{Fabbri:2012yg}.

However, in that paper the model was taken under the assumption of a parity-invariant background, but there is no reason to believe that the gravitational field must display a symmetry that is already known not to pertain to the universe in general; consequently, it is clear that the hypothesis of constraining the action to contain parity-even terms only has to be relaxed: thus, all along the present paper, we shall consider the dynamical action to have all the parity-even but also all the parity-odd contributions as well. This enlarged dynamical action contemplates more terms and it will therefore have a more complicated structure; but even in this case the constraint of least-order derivative will still keep the dynamical action simple enough to try its investigation in the most general instance. Studying the most general of such actions is what we will do in the present article.

In the construction of the most general least-order derivative dynamical action with both parity-even and parity-odd terms we will witness the enlargement of the results presented in \cite{Fabbri:2012yg}, though partial generalizations have already been conducted. For instance, the earliest paper in which all possible terms accounting for both parity-even and parity-odd contributions for both torsion and curvature are considered is reference \cite{Hojman:1980kv}; although in this paper the authors do take into account the most general torsional-gravitational part of the action, nevertheless they proceed dismissing any spinorial contribution whatsoever: because in least-order derivative actions the torsion tensor is algebraically related to the spin of the spinor fields, then there can be no torsion when spinors are absent, and the gravitational action written in terms of purely metric curvatures reduces to the Einstein action alone since the Holst action becomes a surface term vanishing in any classical treatment. As far as we know, Perez and Rovelli have been the first to point out that works wishing to investigate at the classical level the effects of the Holst action cannot neglect torsion, and so spinors must be accounted \cite{Perez:2005pm}; in their work Perez and Rovelli considered Dirac spinor fields, but if one allows both parity-even and parity-odd terms for the torsional-gravitational sector then these terms should be allowed also for the Dirac spinor field, and consequently parity-odd beside parity-even terms for the Dirac action were added first in \cite{Freidel:2005sn} and later in \cite{Mercuri:2006um}; these two generalizations were somewhat complementary, and only when they are both considered one reaches the most general Dirac action as first obtained in \cite{Alexandrov:2008iy}: but nevertheless, even though in this last paper the material action is the most general indeed, the torsional-gravitational action is not the most general action that was already discussed in the above-mentioned work \cite{Hojman:1980kv}, and only when they are both considered one obtains the most general action of all, as we intend to do here. The present paper is therefore written in the spirit of recovering the outline of \cite{Fabbri:2012yg} but for the most general circumstance, the one for which the background is given by the most general torsional-gravitational spacetime as in \cite{Hojman:1980kv} filled with the most general Dirac spinorial matter as in \cite{Alexandrov:2008iy}: after the most general action is written down, we will proceed to obtain the torsion-spin density field equations, which will be algebraic and therefore we may employ them to substitute torsion with the spinor fields obtaining the effective action; the study of this effective action will reveal that all information about the parity-oddness will be lost. That is, we will demonstrate that even starting from the most complete inventory of all parity-odd and parity-even terms, both in the torsional-gravitational and in the material action, eventually there will remain no trace of the parity-oddness but only the usual parity-evenness, in any of the dynamical or self-interacting terms, in the final effective action.
\section{Least-Order Torsion-Gravity for Chiral-Spinor Fields}
\subsection{The Kinematic Construction}
In the present paper, the geometry we will employ is based on the $(1\!+\!3)$-dimensional space-time. The most general connection $\Gamma^{\alpha}_{\mu\nu}$ defines the most general covariant derivative $D_{\mu}$ and such a connection is not symmetric in the two lower indices, that is its antisymmetric part in those indices $\Gamma^{\alpha}_{\mu\nu}\!-\!\Gamma^{\alpha}_{\nu\mu}\!=\! Q^{\alpha}_{\phantom{\alpha}\mu\nu}$ is not zero and it turns out to be a tensor, called Cartan torsion tensor; the two metric tensors $g_{\mu\nu}$ and $g^{\mu\nu}$ symmetric and one the inverse of the other are introduced as usual: the connection and the metric are said to be compatible when the metric-compatibility condition $D_{\mu}g_{\alpha\beta}\!=\!0$ holds. The metric tensor also allows the process of raising-lowering tensorial indices, and after that this process is defined it is possible to build the contorsion tensor given by the expression
\begin{eqnarray}
&K^{\alpha}_{\phantom{\alpha}\mu\nu}=\frac{1}{2}(Q^{\alpha}_{\phantom{\alpha}\mu\nu}
+Q_{\mu\nu}^{\phantom{\mu\nu}\alpha}+Q_{\nu\mu}^{\phantom{\nu\mu}\alpha})
\end{eqnarray}
and the trace vector $Q_{\rho}\!=\!Q^{\alpha}_{\phantom{\alpha}\alpha\rho}\!\equiv\!K_{\rho\alpha}^{\phantom{\rho\alpha}\alpha}$ and we have that
\begin{eqnarray}
&\Gamma^{\alpha}_{\beta\mu}=\Lambda^{\alpha}_{\beta\mu}+K^{\alpha}_{\phantom{\alpha}\beta\mu}
\label{connection}
\end{eqnarray}
is the most general connection, in terms of contorsion and the Levi-Civita connection $\Lambda^{\alpha}_{\beta\mu}$ as usual; the most general connection defines the most general covariant derivative given by $D_{\mu}$ in the same way in which the torsionless Levi-Civita connection defines the torsionless covariant derivative $\nabla_{\mu}$ as it is usual: of course, we have that both conditions $D_{\mu}g_{\alpha\beta}\!=\!\nabla_{\mu}g_{\alpha\beta}\!=\!0$ do hold. Given the metric tensor it is also possible to define the completely antisymmetric Levi-Civita pseudo-tensor $\varepsilon_{\alpha\nu\sigma\iota}$ in terms of which we may proceed to define the axial vector dual of the completely antisymmetric part of torsion or contorsion tensors $W_{\alpha}\!=\!\varepsilon_{\rho\mu\nu\alpha}Q^{\rho\mu\nu} 
\!=\!2\varepsilon_{\rho\mu\nu\alpha}K^{\rho\mu\nu}$ as clear; we then have that $D_{\mu}\varepsilon_{\pi\nu\sigma\iota}\!=\!\nabla_{\mu}\varepsilon_{\pi\nu\sigma\iota}\!=\!0$ as it may be checked with a very straightforward calculation. From the above connection (\ref{connection}) we define another tensor according to
\begin{eqnarray}
&G^{\rho}_{\phantom{\rho}\xi\mu\nu}
=\partial_{\mu}\Gamma^{\rho}_{\xi\nu}-\partial_{\nu}\Gamma^{\rho}_{\xi\mu}
+\Gamma^{\rho}_{\sigma\mu}\Gamma^{\sigma}_{\xi\nu}
-\Gamma^{\rho}_{\sigma\nu}\Gamma^{\sigma}_{\xi\mu}
\label{curvature}
\end{eqnarray}
and from the Levi-Civita connection we define the analogous torsionless tensor given by the similar expression
\begin{eqnarray}
&R^{\rho}_{\phantom{\rho}\xi\mu\nu}
=\partial_{\mu}\Lambda^{\rho}_{\xi\nu}-\partial_{\nu}\Lambda^{\rho}_{\xi\mu}
+\Lambda^{\rho}_{\sigma\mu}\Lambda^{\sigma}_{\xi\nu}
-\Lambda^{\rho}_{\sigma\nu}\Lambda^{\sigma}_{\xi\mu}
\label{metriccurvature}
\end{eqnarray}
such that we have the decomposition
\begin{eqnarray}
\nonumber
&G^{\rho}_{\phantom{\rho}\xi\mu\nu}=R^{\rho}_{\phantom{\rho}\xi\mu\nu}
+\nabla_{\mu}K^{\rho}_{\phantom{\rho}\xi\nu}-\nabla_{\nu}K^{\rho}_{\phantom{\rho}\xi\mu}+\\
&+K^{\rho}_{\phantom{\rho}\sigma\mu}K^{\sigma}_{\phantom{\sigma}\xi\nu}
-K^{\rho}_{\phantom{\rho}\sigma\nu}K^{\sigma}_{\phantom{\sigma}\xi\mu}
\label{separation}
\end{eqnarray}
antisymmetric in the first and second pair of indices, with a single contraction $G^{\rho}_{\phantom{\rho}\mu\rho\nu}\!=\!G_{\mu\nu}$ with contraction given by $G_{\eta\nu}g^{\eta\nu}\!=\!G$ and also $R^{\rho}_{\phantom{\rho}\mu\rho\nu}\!=\!R_{\mu\nu}$ with contraction given by $R_{\eta\nu}g^{\eta\nu}\!=\!R$ respectively called Riemann tensor, Ricci tensor and Ricci scalar, and torsionless Riemann tensor, torsionless Ricci tensor and torsionless Ricci scalar. In Lorentz formalism, the metric is decomposed according to $g_{\alpha\nu}=e_{\alpha}^{p}e_{\nu}^{i} \eta_{pi}$ in terms of the tetrads $e_{\alpha}^{i}$ and the constant metric $\eta_{ij}$ defined to have Minkowskian structure; the connection (\ref{connection}) can be transformed into the spin-connection according to the form
\begin{eqnarray}
&\omega^{i}_{\phantom{i}p\alpha}=
e^{i}_{\sigma}(\Gamma^{\sigma}_{\rho\alpha}e^{\rho}_{p}+\partial_{\alpha}e^{\sigma}_{p})
\label{spin-connection}
\end{eqnarray} 
such that $\omega^{ip}_{\phantom{ip}\alpha}\!=\!-\omega^{pi}_{\phantom{pi}\alpha}$ for the spin-connection, in terms of which the spin-covariant derivative is $D_{i}$ and the torsion tensor can be written according to the expression
\begin{eqnarray}
&-Q^{k}_{\alpha\rho}=\partial_{\alpha}e_{\rho}^{k}-\partial_{\rho}e_{\alpha}^{k}
+\omega^{k}_{\phantom{i}p\alpha}e_{\rho}^{p}-\omega^{k}_{\phantom{i}p\rho}e_{\alpha}^{p}
\label{translation}
\end{eqnarray} 
so that it is possible to decompose the spin-connection down to the torsionless spin-connection and torsional contributions: the relationship (\ref{spin-connection}) and the antisymmetry of the spin-connection are equivalent to the pair of conditions $D_{\mu}e_{i}^{\rho}\!=\!0$ and $D_{\mu}\eta_{ij}\!=\!0$ respectively. We have that it is possible from (\ref{spin-connection}) to write the analogous form
\begin{eqnarray}
&G^{i}_{\phantom{i}j\mu\nu}
=\partial_{\mu}\omega^{i}_{j\nu}-\partial_{\nu}\omega^{i}_{j\mu}
+\omega^{i}_{p\mu}\omega^{p}_{j\nu}-\omega^{i}_{p\nu}\omega^{p}_{j\mu}
\label{rotation}
\end{eqnarray}
for which we have that 
\begin{eqnarray}
&G^{i}_{\phantom{i}j\mu\nu}\!=\!G^{\rho}_{\phantom{\rho}\sigma\mu\nu} e^{\sigma}_{j}e^{i}_{\rho}
\end{eqnarray}
as it should be for consistency, since such an object is a tensor and therefore the passage from general coordinate to Lorentz formalisms must be given in terms of a simple index renaming. The reason for which it is useful to pass from general coordinate to Lorentz formalism is that torsion and curvature as in (\ref{translation}-\ref{rotation}) are respectively seen as strength of the tetrads and spin-connection, when these are interpreted as potentials obtained by gauging the roto-translations of the Poincar\'{e} group; but more importantly, such a Lorentz formalism is essential because Lorentz-valued indices are subject to tensorial transformations belonging to the Lorentz group, which can be written explicitly and therefore given in terms of other, different representations. Of particular interest for us are the complex representations of the Lorentz group.

Here, the complex representations of the Lorentz group will be in $\frac{1}{2}$-spin representation. Such a representation can be achieved through the introduction of the $\boldsymbol{\gamma}_{a}$ matrices such that $\{\boldsymbol{\gamma}_{a},\boldsymbol{\gamma}_{b}\}
\!=\!2\boldsymbol{\mathbb{I}}\eta_{ab}$ from which one may define the matrices $\boldsymbol{\sigma}_{ab}\!=\!\frac{1}{4}[\boldsymbol{\gamma}_{a},\boldsymbol{\gamma}_{b}]$ as the infinitesimal complex generators of the complex Lorentz transformation for the $\frac{1}{2}$-spin spinorial field; these matrices also verify the relations $\{\boldsymbol{\gamma}_{a},
\boldsymbol{\sigma}_{bc}\}\!=\!i\varepsilon_{abcd} \boldsymbol{\pi}\boldsymbol{\gamma}^{d}$ implicitly defining the $\boldsymbol{\pi}$ matrix that will be used to define the left-handed and the right-handed chiral projectors, which will become important to define left-handed and right-handed chiral semi-spinor fields. From (\ref{spin-connection}) it is possible to define the spinorial-connection according to the form
\begin{eqnarray}
&\boldsymbol{\Omega}_{\rho}
=\frac{1}{2}\omega^{ij}_{\phantom{ij}\rho}\boldsymbol{\sigma}_{ij}
\label{spinorial-connection}
\end{eqnarray} 
defining $\boldsymbol{D}_{\mu}\psi$ as the spinorial covariant derivative acting on the spinorial fields: conditions $\boldsymbol{D}_{\mu}\boldsymbol{\gamma}_{a}\!=\!0$ are valid automatically. We then define the spinorial tensor 
\begin{eqnarray}
&\boldsymbol{G}_{\mu\nu}=\frac{1}{2}G^{ij}_{\phantom{ij}\mu\nu}\boldsymbol{\sigma}_{ij}
\label{spinorial-tensor}
\end{eqnarray} 
in terms of which $[\boldsymbol{D}_{\mu},\boldsymbol{D}_{\nu}]\psi\!=\! Q^{\rho}_{\phantom{\rho}\mu\nu} \boldsymbol{D}_{\rho}\psi\!+\!\boldsymbol{G}_{\mu\nu}\psi$ is the commutator of the spinorial covariant derivatives of the spinor field in the most general form. The tetrads and the spin-connection are essential to encode the spin structure.
\subsection{The Dynamical Action}
Having defined the kinematic quantities, we proceed to the construction of the least-order derivative dynamical action in the most general case with parity-even and parity-odd terms. As in \cite{Hojman:1980kv}, it is easy to see that the inventory of all parity-even and parity-odd contributions for the torsional-gravity background gives nine terms
\begin{eqnarray}
\nonumber
&\mathcal{L}_{\mathrm{g}}=(HQ_{\rho\mu\nu}Q^{\rho}_{\phantom{\rho}\alpha\beta}
\!+\!IQ_{\mu\nu\rho}Q_{\alpha\beta}^{\phantom{\alpha\beta}\rho}
\!+\!JQ_{\rho\mu\nu}Q_{\alpha\beta}^{\phantom{\alpha\beta}\rho}+\\
\nonumber
&+KQ_{\mu}Q_{\nu\alpha\beta}+kG_{\mu\nu\alpha\beta})\varepsilon^{\mu\nu\alpha\beta}+\\
&+(AQ_{\mu}Q^{\mu}\!+\!BQ_{\rho\mu\nu}Q^{\rho\mu\nu}\!+\!CQ_{\rho\mu\nu}Q^{\nu\mu\rho}+G)
\label{geometry}
\end{eqnarray}
where the Newton constant is normalized to unity and the constant $k$ is the Immirzi parameter; by decomposing all curvatures into torsionless curvature plus torsional contributions, we have that this action is equivalent to
\begin{eqnarray}
\nonumber
&\mathcal{L}_{\mathrm{g}}=[KQ_{\mu}Q_{\nu\alpha\beta}\!+\!
JQ_{\rho\mu\nu}Q_{\alpha\beta}^{\phantom{\alpha\beta}\rho}+\\
\nonumber
&+IQ_{\mu\nu\rho}Q_{\alpha\beta}^{\phantom{\alpha\beta}\rho}
\!+\!\frac{1}{2}FQ_{\rho\mu\nu}Q^{\rho}_{\phantom{\rho}\alpha\beta}]
\varepsilon^{\mu\nu\alpha\beta}+\\
&+[XQ_{\mu}Q^{\mu}\!+\!YQ_{\rho\mu\nu}Q^{\rho\mu\nu}
\!+\!ZQ_{\rho\mu\nu}Q^{\nu\mu\rho}]\!+\!R
\label{gravity}
\end{eqnarray}
where we have called $2H\!+\!k\!=\!F$, and having also renamed the constants $A\!-\!1\!=\!X$, 
$B\!+\!\frac{1}{4}\!=\!Y$, $C\!+\!\frac{1}{2}\!=\!Z$, for the sake of simplicity: the Immirzi parameter $k$ is lost in the parity-odd parameter $F$ which, with $I$, $J$, $K$, form the set of $4$ parameters labelling the parity-odd terms, beside the usual $X$, $Y$, $Z$, as the set of $3$ parameters labelling the parity-even terms, amounting to a total of $7$ parameters for the torsional terms, completing the single parity-even curvature term given by the usual gravitational action. 

As in \cite{Alexandrov:2008iy}, it is not difficult to prove that all parity-even and all parity-odd contributions are accounted for the Dirac spinorial matter by the action given according to
\begin{eqnarray}
\nonumber
&\mathcal{L}_{\mathrm{m}}=\frac{i}{2}[\overline{\psi}\boldsymbol{\gamma}^{\mu}
(a\!+\!ib\boldsymbol{\pi})\boldsymbol{D}_{\mu}\psi
\!-\!\boldsymbol{D}_{\mu}\overline{\psi}\boldsymbol{\gamma}^{\mu}
(a^{\ast}\!-\!ib^{\ast}\boldsymbol{\pi})\psi]-\\
&-\mu\overline{\psi}\psi\!-\!\beta i\overline{\psi}\boldsymbol{\pi}\psi
\label{matter}
\end{eqnarray}
where $a$ and $b$ are complex, while $\mu$ and $\beta$ must be real, to ensure the reality of the overall action; as above we may decompose all spinorial covariant derivatives into torsionless spinorial covariant derivatives plus torsional contributions, but now we may additionally split the spinors into their left-handed $\frac{1}{2}(\mathbb{I}\!-\!\boldsymbol{\pi})\psi\!=\!\psi_{L}$ and the complementary right-handed $\frac{1}{2}(\mathbb{I}\!+\!\boldsymbol{\pi})\psi\!=\!\psi_{R}$ semi-spinor fields.

When this is done the total action $\mathcal{L}\!=\!\mathcal{L}_{\mathrm{g}}\!-\!\mathcal{L}_{\mathrm{m}}$ becomes
\begin{eqnarray}
\nonumber
&\mathcal{L}=[KQ_{\mu}Q_{\nu\alpha\beta}\!+\!
JQ_{\rho\mu\nu}Q_{\alpha\beta}^{\phantom{\alpha\beta}\rho}+\\
\nonumber
&+IQ_{\mu\nu\rho}Q_{\alpha\beta}^{\phantom{\alpha\beta}\rho}
\!+\!\frac{1}{2}FQ_{\rho\mu\nu}Q^{\rho}_{\phantom{\rho}\alpha\beta}]
\varepsilon^{\mu\nu\alpha\beta}+\\
\nonumber
&+[XQ_{\mu}Q^{\mu}\!+\!YQ_{\rho\mu\nu}Q^{\rho\mu\nu}
\!+\!ZQ_{\rho\mu\nu}Q^{\nu\mu\rho}]\!+R\!-\\
\nonumber
&-i[\frac{1}{2}(a\!+\!a^{\ast}\!)\!-\!\frac{i}{2}(b\!-\!b^{\ast}\!)]
\overline{\psi}_{L}\boldsymbol{\gamma}^{\mu}\boldsymbol{\nabla}_{\mu}\psi_{L}-\\
\nonumber
&-i[\frac{1}{2}(a\!+\!a^{\ast}\!)\!+\!\frac{i}{2}(b\!-\!b^{\ast}\!)]
\overline{\psi}_{R}\boldsymbol{\gamma}^{\mu}\boldsymbol{\nabla}_{\mu}\psi_{R}+\\
\nonumber
&+\frac{1}{8}\left[\frac{1}{2}(a\!+\!a^{\ast}\!)\!-\!\frac{i}{2}(b\!-\!b^{\ast}\!)\right]
Q_{\mu\alpha\beta}\varepsilon^{\mu\alpha\beta\rho}
\overline{\psi}_{L}\boldsymbol{\gamma}_{\rho}\psi_{L}-\\
\nonumber
&-\frac{1}{8}\left[\frac{1}{2}(a\!+\!a^{\ast}\!)\!+\!\frac{i}{2}(b\!-\!b^{\ast}\!)\right]
Q_{\mu\alpha\beta}\varepsilon^{\mu\alpha\beta\rho}
\overline{\psi}_{R}\boldsymbol{\gamma}_{\rho}\psi_{R}+\\
\nonumber
&+\frac{i}{2}\left[\frac{1}{2}(a\!-\!a^{\ast}\!)\!-\!\frac{i}{2}(b\!+\!b^{\ast}\!)\right]
Q_{\beta}\overline{\psi}_{L}\boldsymbol{\gamma}^{\beta}\psi_{L}+\\
\nonumber
&+\frac{i}{2}\left[\frac{1}{2}(a\!-\!a^{\ast}\!)\!+\!\frac{i}{2}(b\!+\!b^{\ast}\!)\right]
Q_{\beta}\overline{\psi}_{R}\boldsymbol{\gamma}^{\beta}\psi_{R}+\\
\nonumber
&+\mu(\overline{\psi}_{R}\psi_{L}+\overline{\psi}_{L}\psi_{R})+\\
&+i\beta(\overline{\psi}_{L}\psi_{R}\!-\!\overline{\psi}_{R}\psi_{L})
\label{action}
\end{eqnarray}
where in the left-handed and right-handed kinetic semi-spinorial terms the two real factors can then be renamed like $\frac{1}{2}(a\!+\!a^{\ast}\!-\!ib\!+\!ib^{\ast}\!)\!=\!p$ and 
$\frac{1}{2}(a\!+\!a^{\ast}\!+\!ib\!-\!ib^{\ast}\!)\!=\!q$ as real constants, which can be absorbed into a rescaling of the left-handed and right-handed semi-spinorial fields and a redefinition of all other constants, so that there is no loss of generality in setting $p\!=\!q\!=\!1$ as it is customary.
\section{Torsionally-Induced Self-Interacting Potentials}
\subsection{The Torsion-Spin Algebraic Coupling}
When the action (\ref{action}) is varied with respect to torsion, one gets the field equations that couple the torsion tensor to the spin density tensor of the spinor field as 
\begin{eqnarray}
\nonumber
&K(Q_{\theta}\varepsilon^{\theta\rho\mu\nu}\!+\!\frac{1}{2}g^{\rho[\nu}W^{\mu]})+\\
\nonumber
&+J(Q_{\sigma\alpha}^{\phantom{\sigma\alpha}\rho}\varepsilon^{\sigma\alpha\mu\nu}
\!+\!\frac{1}{2}\varepsilon^{\sigma\alpha\rho[\mu}Q^{\nu]}_{\phantom{\nu}\sigma\alpha})-\\
\nonumber
&-IQ_{\alpha\sigma}^{\phantom{\alpha\sigma}[\nu}\varepsilon^{\mu]\rho\alpha\sigma}
+FQ^{\rho}_{\phantom{\rho}\alpha\sigma}\varepsilon^{\alpha\sigma\mu\nu}+\\
\nonumber
&+Xg^{\rho[\mu}Q^{\nu]}\!+\!2YQ^{\rho\mu\nu}\!+\!Z(Q^{\nu\mu\rho}\!-\!Q^{\mu\nu\rho})=\\
\nonumber
&=-\frac{1}{8}\left[\frac{1}{2}(a\!+\!a^{\ast}\!)\!-\!\frac{i}{2}(b\!-\!b^{\ast}\!)\right]
\varepsilon^{\rho\mu\nu\alpha}\overline{\psi}_{L}\boldsymbol{\gamma}_{\alpha}\psi_{L}+\\
\nonumber
&+\frac{1}{8}\left[\frac{1}{2}(a\!+\!a^{\ast}\!)\!+\!\frac{i}{2}(b\!-\!b^{\ast}\!)\right]
\varepsilon^{\rho\mu\nu\alpha}\overline{\psi}_{R}\boldsymbol{\gamma}_{\alpha}\psi_{R}-\\
\nonumber
&-\frac{i}{4}\left[\frac{1}{2}(a\!-\!a^{\ast}\!)\!-\!\frac{i}{2}(b\!+\!b^{\ast}\!)\right]
g^{\rho[\mu}\overline{\psi}_{L}\boldsymbol{\gamma}^{\nu]}\psi_{L}-\\
&-\frac{i}{4}\left[\frac{1}{2}(a\!-\!a^{\ast}\!)\!+\!\frac{i}{2}(b\!+\!b^{\ast}\!)\right]
g^{\rho[\mu}\overline{\psi}_{R}\boldsymbol{\gamma}^{\nu]}\psi_{R}
\label{Sciama-Kibble}
\end{eqnarray}
where the spin density decomposes in terms of its axial and trace parts only; since torsion is algebraically related to the spin density, it can be decomposed according to
\begin{eqnarray}
&Q_{\rho\mu\nu}\!=\!\frac{1}{3}(g_{\rho\mu}Q_{\nu}\!-\!g_{\rho\nu}Q_{\mu})
\!+\!\frac{1}{6}W^{\beta}\varepsilon_{\beta\rho\mu\nu}
\label{decomposition}
\end{eqnarray}
in terms of the axial vector torsion and the trace vector torsion alone (see appendix \ref{a1}): as both torsion and spin have only the axial and trace parts, (\ref{Sciama-Kibble}) are equivalent to the field equations for the spin axial vector given by
\begin{eqnarray}
\nonumber
&\Delta Q^{\beta}\!+\!\Phi W^{\beta}=\\
\nonumber
&=\frac{3}{8}\left[\frac{1}{2}(a\!+\!a^{\ast}\!)\!-\!\frac{i}{2}(b\!-\!b^{\ast}\!)\right]
\overline{\psi}_{L}\boldsymbol{\gamma}^{\beta}\psi_{L}-\\
&-\frac{3}{8}\left[\frac{1}{2}(a\!+\!a^{\ast}\!)\!+\!\frac{i}{2}(b\!-\!b^{\ast}\!)\right]
\overline{\psi}_{R}\boldsymbol{\gamma}^{\beta}\psi_{R}
\label{Sciama}
\end{eqnarray}
plus the field equations for the spin trace vector as
\begin{eqnarray}
\nonumber
&-\Delta W^{\nu}\!+\!2\Theta Q^{\nu}=\\
\nonumber
&=-\frac{3i}{2}\left[\frac{1}{2}(a\!-\!a^{\ast}\!)
\!-\!\frac{i}{2}(b\!+\!b^{\ast}\!)\right]\overline{\psi}_{L}\boldsymbol{\gamma}^{\nu}\psi_{L}-\\
&-\frac{3i}{2}\left[\frac{1}{2}(a\!-\!a^{\ast}\!)
\!+\!\frac{i}{2}(b\!+\!b^{\ast}\!)\right]\overline{\psi}_{R}\boldsymbol{\gamma}^{\nu}\psi_{R}
\label{Kibble}
\end{eqnarray}
having renamed the parameters according to
\begin{eqnarray}
&3K\!-\!J\!+\!2I\!-\!2F\!=\!\Delta\\
&3X\!+\!2Y\!+\!Z\!=\!\Theta\\
&Y\!-\!Z\!=\!\Phi
\end{eqnarray}
for the parity-odd and for both parity-even coefficients, for simplicity. These relationships can be inverted in order to get field equations for the axial vector torsion as
\begin{eqnarray}
&\frac{16}{3}[2\Theta\Phi+\Delta^{2}]Q^{\nu}
=\upsilon\overline{\psi}_{L}\boldsymbol{\gamma}^{\nu}\psi_{L}
-\omega\overline{\psi}_{R}\boldsymbol{\gamma}^{\nu}\psi_{R}
\label{trace}
\end{eqnarray}
and field equations for the trace vector torsion as
\begin{eqnarray}
&\frac{8}{3}[2\Theta\Phi+\Delta^{2}]W^{\nu}
=\zeta\overline{\psi}_{L}\boldsymbol{\gamma}^{\nu}\psi_{L}
-\xi\overline{\psi}_{R}\boldsymbol{\gamma}^{\nu}\psi_{R}
\label{axial}
\end{eqnarray}
where once again we have renamed the parameters as
\begin{eqnarray}
&(\Delta\!-\!4i\Phi)(a\!-\!ib)\!+\!(\Delta\!-\!4i\Phi)^{\ast}(a\!-\!ib)^{\ast}\!=\!\upsilon
\label{1}\\
&(\Delta\!+\!4i\Phi)(a\!+\!ib)\!+\!(\Delta\!+\!4i\Phi)^{\ast}(a\!+\!ib)^{\ast}\!=\!\omega
\label{2}\\
&(\Theta\!+\!2i\Delta)(a\!-\!ib)\!+\!(\Theta\!+\!2i\Delta)^{\ast}(a\!-\!ib)^{\ast}\!=\!\zeta
\label{3}\\
&(\Theta\!-\!2i\Delta)(a\!+\!ib)\!+\!(\Theta\!-\!2i\Delta)^{\ast}(a\!+\!ib)^{\ast}\!=\!\xi
\label{4}
\end{eqnarray}
so to simplify the notation in what follows: the explicit expression for torsion is given by (\ref{decomposition}) with (\ref{trace}-\ref{axial}) as
\begin{eqnarray}
\nonumber
&16[2\Theta\Phi+\Delta^{2}]Q_{\rho\mu\nu}=\\
\nonumber
&=\upsilon g_{\rho[\mu}\overline{\psi}_{L}\boldsymbol{\gamma}_{\nu]}\psi_{L}
-\omega g_{\rho[\mu}\overline{\psi}_{R}\boldsymbol{\gamma}_{\nu]}\psi_{R}+\\
&+\zeta \varepsilon_{\beta\rho\mu\nu}\overline{\psi}_{L}\boldsymbol{\gamma}^{\beta}\psi_{L}
-\xi \varepsilon_{\beta\rho\mu\nu}\overline{\psi}_{R}\boldsymbol{\gamma}^{\beta}\psi_{R}
\label{torsion}
\end{eqnarray}
with condition $2\Theta\Phi\!+\!\Delta^{2}\!\neq\!0$ to ensure torsion is inverted and such that $\upsilon\!=\!-\omega$ and $\zeta\!=\!\xi$ in order for the inverted torsion to be parity-even as it should. As the torsion-spin density field equations are algebraic, torsion can be substituted in terms of the spinor in the original action.

After the substitution is done, we have that all squares of torsion and the products between torsion and the current density of the spinor fields become either squares of currents $\overline{\psi}_{R}\boldsymbol{\gamma}^{\nu}\psi_{R} \overline{\psi}_{R}\boldsymbol{\gamma}_{\nu}\psi_{R}$ and $\overline{\psi}_{L}\boldsymbol{\gamma}^{\nu}\psi_{L} \overline{\psi}_{L}\boldsymbol{\gamma}_{\nu}\psi_{L}$ or scalar products of currents $\overline{\psi}_{L}\boldsymbol{\gamma}^{\nu}\psi_{L}
\overline{\psi}_{R}\boldsymbol{\gamma}_{\nu}\psi_{R}$ as it is quite easy to check by looking at all possible combinations; through the Fierz identities (see appendix \ref{a2}), we may acknowledge that one has $\overline{\psi}_{R}\boldsymbol{\gamma}_{\nu}\psi_{R}
\overline{\psi}_{R}\boldsymbol{\gamma}^{\nu}\psi_{R}\!=\!0$ together with the analogous $\overline{\psi}_{L}\boldsymbol{\gamma}_{\nu}\psi_{L}
\overline{\psi}_{L}\boldsymbol{\gamma}^{\nu}\psi_{L}\!=\!0$ implying that the squared currents vanish leaving the scalar product of the two currents as $\overline{\psi}_{R}\boldsymbol{\gamma}_{\nu}\psi_{R}
\overline{\psi}_{L}\boldsymbol{\gamma}^{\nu}\psi_{L}$ alone in the effective action
\begin{eqnarray}
\nonumber
&\mathcal{L}\!=\!R\!
-\!i\overline{\psi}_{L}\boldsymbol{\gamma}^{\mu}\boldsymbol{\nabla}_{\mu}\psi_{L}
-i\overline{\psi}_{R}\boldsymbol{\gamma}^{\mu}\boldsymbol{\nabla}_{\mu}\psi_{R}-\\
\nonumber
&-\frac{3}{32}\lambda\ \overline{\psi}_{L}\boldsymbol{\gamma}^{\nu}\psi_{L}\
\overline{\psi}_{R}\boldsymbol{\gamma}_{\nu}\psi_{R}+\\
\nonumber
&+\mu(\overline{\psi}_{R}\psi_{L}+\overline{\psi}_{L}\psi_{R})+\\
&+i\beta(\overline{\psi}_{L}\psi_{R}\!-\!\overline{\psi}_{R}\psi_{L})
\label{actional}
\end{eqnarray}
in terms of the parameter $\lambda$ which is a real constant: it is not necessary to give the explicit expression in terms of which all the previous parameters combined in order to give the new parameter since the only concept one need retain is the fact that those parameters can be collected within a single parameter $\lambda$ and that this single parameter is a real coefficient. It is important to stress that the invertibility condition given by $2\Theta\Phi\!+\!\Delta^{2}\!\neq\!0$ is assumed.

If such a condition is instead not valid, then we have that the constraint $2\Theta\Phi\!+\!
\Delta^{2}\!=\!0$ implies that the torsion-spin density field equations (\ref{trace}-\ref{axial}) cannot be used to invert torsion but they simply yield the constraint given in the form $\upsilon(\overline{\psi}_{L}\boldsymbol{\gamma}^{\nu}\psi_{L}\!+\!
\overline{\psi}_{R}\boldsymbol{\gamma}^{\nu}\psi_{R})\!=\!0$ together with the complementary $\xi(\overline{\psi}_{L}\boldsymbol{\gamma}^{\nu}\psi_{L}\!-\!
\overline{\psi}_{R}\boldsymbol{\gamma}^{\nu}\psi_{R})\!=\!0$ which in turn can be worked out to see that $\upsilon\!=\!\xi\!=\!0$ since otherwise the spinor field would vanish identically: when the pair of constraints $\upsilon\!=\!\xi\!=\!0$ is used in (\ref{1}-\ref{4}) it is straightforward to prove that we also get the pair of constraints given by $\Theta\!=\!-\Delta(b\!+\!b^{\ast}\!)$ and $\Delta\!=\!2\Phi(b\!+\!b^{\ast}\!)$ together with the additional condition $a\!=\!a^{\ast}$ in terms of which the torsion-spin field equations (\ref{Sciama}-\ref{Kibble}) become
\begin{eqnarray}
\nonumber
&\Delta Q^{\beta}\!+\!\Phi W^{\beta}=\\
&=\frac{3}{8}(\overline{\psi}_{L}\boldsymbol{\gamma}^{\beta}\psi_{L}
\!-\!\overline{\psi}_{R}\boldsymbol{\gamma}^{\beta}\psi_{R})
\label{Sciamared}
\end{eqnarray}
together with the complementary
\begin{eqnarray}
\nonumber
&-\Delta W^{\nu}\!+\!2\Theta Q^{\nu}=\\
&=-\frac{3}{4}(b\!+\!b^{\ast}\!)(\overline{\psi}_{L}\boldsymbol{\gamma}^{\nu}\psi_{L}
\!-\!\overline{\psi}_{R}\boldsymbol{\gamma}^{\nu}\psi_{R})
\label{Kibblered}
\end{eqnarray}
as it is easy to check; the two are both equivalent to the single field equation for the two vectorial parts of torsion
\begin{eqnarray}
&2\Phi(b\!+\!b^{\ast}\!)Q^{\beta}\!+\!\Phi W^{\beta}
=\frac{3}{8}(\overline{\psi}_{L}\boldsymbol{\gamma}^{\beta}\psi_{L}
\!-\!\overline{\psi}_{R}\boldsymbol{\gamma}^{\beta}\psi_{R})
\end{eqnarray}
which cannot be inverted to write the two vectorial parts of torsion separately. However, it is possible to write one vector in terms of the other plus the spinorial current, and substituting this back into the original action gives the result for which all terms depending on the remaining vectorial part of torsion disappear, leaving only the squares of the spinorial currents; these can be Fierz rearranged, leading again to the effective action given by
\begin{eqnarray}
\nonumber
&\mathcal{L}\!=\!R\!
-\!i\overline{\psi}_{L}\boldsymbol{\gamma}^{\mu}\boldsymbol{\nabla}_{\mu}\psi_{L}
-i\overline{\psi}_{R}\boldsymbol{\gamma}^{\mu}\boldsymbol{\nabla}_{\mu}\psi_{R}-\\
\nonumber
&-\frac{3}{32}\lambda\ \overline{\psi}_{L}\boldsymbol{\gamma}^{\nu}\psi_{L}\
\overline{\psi}_{R}\boldsymbol{\gamma}_{\nu}\psi_{R}+\\
\nonumber
&+\mu(\overline{\psi}_{R}\psi_{L}+\overline{\psi}_{L}\psi_{R})+\\
&+i\beta(\overline{\psi}_{L}\psi_{R}\!-\!\overline{\psi}_{R}\psi_{L})
\label{actionalred}
\end{eqnarray}
in terms of the parameter $\lambda$ which is a real constant, and as one may expect it is exactly the same parameter that has been introduced above. However, once again in this derivation we have taken $\Delta\!\neq\!0$ to be valid throughout.

If this does not hold then $\Delta\!=\!0$ implies that there is another constraint given by $\Theta=0$ and so the torsion-spin density field equations (\ref{Sciamared}-\ref{Kibblered}) yield the additional condition $b\!=\!-b^{\ast}$ reducing to the single equation
\begin{eqnarray}
&\Phi W^{\beta}\!=\!\frac{3}{8}(\overline{\psi}_{L}\boldsymbol{\gamma}^{\beta}\psi_{L}
\!-\!\overline{\psi}_{R}\boldsymbol{\gamma}^{\beta}\psi_{R})
\end{eqnarray}
from which the trace vector part of torsion cannot be determined, but there is no need for that since such a trace vector part of torsion has also disappeared form the original action; thus when the axial vector torsion is substituted back into the original action, and the Fierz identities are taken, we get once again the effective action
\begin{eqnarray}
\nonumber
&\mathcal{L}\!=\!R\!
-\!i\overline{\psi}_{L}\boldsymbol{\gamma}^{\mu}\boldsymbol{\nabla}_{\mu}\psi_{L}
-i\overline{\psi}_{R}\boldsymbol{\gamma}^{\mu}\boldsymbol{\nabla}_{\mu}\psi_{R}-\\
\nonumber
&-\frac{3}{32}\lambda\ \overline{\psi}_{L}\boldsymbol{\gamma}^{\nu}\psi_{L}\
\overline{\psi}_{R}\boldsymbol{\gamma}_{\nu}\psi_{R}+\\
\nonumber
&+\mu(\overline{\psi}_{R}\psi_{L}+\overline{\psi}_{L}\psi_{R})+\\
&+i\beta(\overline{\psi}_{L}\psi_{R}\!-\!\overline{\psi}_{R}\psi_{L})
\label{actionalredalt}
\end{eqnarray}
in terms of the parameter $\lambda$ which is a real constant and it cannot be singular as it is already discussed above.
\section{Parity Conservation}
The effective actions (\ref{actional}-\ref{actionalred}-\ref{actionalredalt}) are all formally the same, so it is without losing generality that we will consider them to be written as a single action with a unique constant, and they can further be manipulated in order for the non-linear potential to be rearranged as
\begin{eqnarray}
\nonumber
&\mathcal{L}\!=\!R\!-\!
i\overline{\psi}_{L}\boldsymbol{\gamma}^{\mu}\boldsymbol{\nabla}_{\mu}\psi_{L}
-i\overline{\psi}_{R}\boldsymbol{\gamma}^{\mu}\boldsymbol{\nabla}_{\mu}\psi_{R}-\\
\nonumber
&-\frac{3\lambda}{16}\overline{\psi}_{R}\psi_{L}\overline{\psi}_{L}\psi_{R}
\!+\!\mu(\overline{\psi}_{R}\psi_{L}\!+\!\overline{\psi}_{L}\psi_{R})+\\
&+i\beta(\overline{\psi}_{L}\psi_{R}\!-\!\overline{\psi}_{R}\psi_{L})
\end{eqnarray}
in which any parity-oddness disappeared leaving only parity-even contributions in the torsionally-induced self-spinorial interactions: although we have allowed parity-odd and parity-even terms within the most general action possible, we have obtained what we would have had had we allowed only parity-even terms; this is due to the constrained structure of the Dirac matter fields and the limited number of ways in which its components can be rearranged by Fierz identities. As a consequence, it is the very character of the matter field that ultimately enables the effective action to ensure parity conservation.
\section*{Conclusion}
In this paper, we have started from the most comprehensive inventory of both parity-odd and parity-even terms for the torsional completion of gravity in an underlying background filled with Dirac matter fields, writing the most general least-order derivative action, and we have derived some consequences; in particular, we have obtained the torsion-spin coupling field equations, which we have decomposed in terms of the axial vector and trace vector parts, inverting them in order to write these in terms of the left-handed and right-handed semi-spinorial currents; due to their algebraic character, we have substituted the torsion tensor in terms of the spin density currents back into the original action, getting an effective action that has been investigated: we have seen that the torsionally-induced self-spinorial interactions were parity conserving nonetheless. This result means that despite generalizing the dynamics the non-linear potentials come always down to a parity conserving one; that is, although we may renounce to parity in the action, we will get parity in the effective action anyway. It may appear that parity possesses the property of being stable under breaking, in some sort of parity preserving mechanism.

What we have done can be summarized by saying that following a spirit of generalization, we have obtained the most general least-order derivative torsional completion of gravity with Dirac matter having both parity-odd and parity-even terms by combining the results of references \cite{Hojman:1980kv} and \cite{Alexandrov:2008iy}, although we have obtained that the entire parity-odd contribution disappeared leaving only the parity-even terms as if we were to start from the general action discussed in \cite{Fabbri:2012yg}: notice that when Fierz rearrangements are taken into account it is possible to prove that also the effective action in \cite{Alexandrov:2008iy} is reconducted to the parity-even form that was presented in \cite{Fabbri:2012yg}; a very similar situation happened precisely in reference \cite{Fabbri:2012yg}, where we wrote the most general least-order derivative torsional completion of gravity with Dirac matter including beside the square of the axial torsion also the square of the other two torsional parts, but we saw that these two latter contributions disappeared leaving only the axial torsion contribution as if working under the constraint of the complete antisymmetry of torsion. Notice that the two reductions have the same reason, that is no matter how general the generalization is, either with or without a definite parity, or with or without squared of the non-completely antisymmetric parts of torsion, all parity-odd and non-axial torsion contributions will disappear, leaving only parity-even axial-vector torsion contributions alone in the final form of the effective action. It is not possible to generalize the form of non-linear potentials due to the constrained structure of the Dirac material field and the correspondingly limited number of ways in which its components can be rearranged by employing Fierz identities.

As parity conservation is a feature of the Einstein gravitational action then any parity violating term can only be present within the Dirac matter field alone, and although we have seen that parity conservation is a feature of the non-linear potential, it is nevertheless possible for the Dirac matter field to have a mass term with an imaginary part proportional to the $\beta$ parameter \cite{Fabbri:2010rw}, breaking parity, though in an arbitrary way; despite in principle such a term is possible, we have nonetheless no insight on its physical effects for now. Physical effects of the parity-violating Holst action are instead supposed to be relevant in quantum treatments \cite{Holst:1995pc}, despite the fact that the Holst action in absence of torsion is just a surface term that can be wholly neglected; conversely, here we did not neglect torsion, though all calculations were done without field quantization. Admittedly, we find rather intriguing that the presence of torsion on the one hand and the hypothesis of field quantization on the other hand have the analogous effect of rendering relevant terms that otherwise would not have been relevant whatsoever.

If there really were a sort of deeper connection between the spacetime torsion and the protocol of field quantization, that would be quite curious.
\appendix \footnotesize
\section{Proof of the torsional constraints}
\label{a1}
In this paper we have stated that starting from the torsion-spin coupling field equations (\ref{Sciama-Kibble}) it was possible to prove that only the axial vector and trace vector torsion-spin coupling field equations given by (\ref{Sciama}-\ref{Kibble}) were relevant: now it is the place to prove it, and to begin let us first of all consider the full field equations (\ref{Sciama-Kibble})
\begin{eqnarray}
\nonumber
&K(Q_{\theta}\varepsilon^{\theta\rho\mu\nu}\!+\!\frac{1}{2}g^{\rho[\nu}W^{\mu]})
\!+\!Xg^{\rho[\mu}Q^{\nu]}+\\
\nonumber
&+J(Q_{\sigma\alpha}^{\phantom{\sigma\alpha}\rho}\varepsilon^{\sigma\alpha\mu\nu}
\!+\!\frac{1}{2}\varepsilon^{\sigma\alpha\rho[\mu}Q^{\nu]}_{\phantom{\nu}\sigma\alpha})
\!+\!2YQ^{\rho\mu\nu}-\\
\nonumber
&-IQ_{\alpha\sigma}^{\phantom{\alpha\sigma}[\nu}\varepsilon^{\mu]\rho\alpha\sigma}
+FQ^{\rho}_{\phantom{\rho}\alpha\sigma}\varepsilon^{\alpha\sigma\mu\nu}
\!+\!Z(Q^{\nu\mu\rho}\!-\!Q^{\mu\nu\rho})=\\
\nonumber
&=-\frac{1}{8}\varepsilon^{\rho\mu\nu\alpha}\overline{\psi}_{L}\boldsymbol{\gamma}_{\alpha}\psi_{L}
\left[\frac{1}{2}(a\!+\!a^{\ast}\!)\!-\!\frac{i}{2}(b\!-\!b^{\ast}\!)\right]+\\
\nonumber
&+\frac{1}{8}\varepsilon^{\rho\mu\nu\alpha}\overline{\psi}_{R}\boldsymbol{\gamma}_{\alpha}\psi_{R}
\left[\frac{1}{2}(a\!+\!a^{\ast}\!)\!+\!\frac{i}{2}(b\!-\!b^{\ast}\!)\right]-\\
\nonumber
&-\frac{i}{4}g^{\rho[\mu}\overline{\psi}_{L}\boldsymbol{\gamma}^{\nu]}\psi_{L}
\left[\frac{1}{2}(a\!-\!a^{\ast}\!)\!-\!\frac{i}{2}(b\!+\!b^{\ast}\!)\right]\\
&-\frac{i}{4}g^{\rho[\mu}\overline{\psi}_{R}\boldsymbol{\gamma}^{\nu]}\psi_{R}
\left[\frac{1}{2}(a\!-\!a^{\ast}\!)\!+\!\frac{i}{2}(b\!+\!b^{\ast}\!)\right]
\label{SK}
\end{eqnarray}
taking their completely antisymmetric part and their trace, in order to get the axial vector part given according to the expression
\begin{eqnarray}
\nonumber
&\Delta Q^{\beta}\!+\!\Phi W^{\beta}=\\
\nonumber
&=\frac{3}{8}\left[\frac{1}{2}(a\!+\!a^{\ast}\!)\!-\!\frac{i}{2}(b\!-\!b^{\ast}\!)\right]
\overline{\psi}_{L}\boldsymbol{\gamma}^{\beta}\psi_{L}-\\
&-\frac{3}{8}\left[\frac{1}{2}(a\!+\!a^{\ast}\!)\!+\!\frac{i}{2}(b\!-\!b^{\ast}\!)\right]
\overline{\psi}_{R}\boldsymbol{\gamma}^{\beta}\psi_{R}
\label{S}
\end{eqnarray}
and their trace vector part given by
\begin{eqnarray}
\nonumber
&-\Delta W^{\nu}\!+\!2\Theta Q^{\nu}=\\
\nonumber
&=-\frac{3i}{2}\left[\frac{1}{2}(a\!-\!a^{\ast}\!)
\!-\!\frac{i}{2}(b\!+\!b^{\ast}\!)\right]\overline{\psi}_{L}\boldsymbol{\gamma}^{\nu}\psi_{L}-\\
&-\frac{3i}{2}\left[\frac{1}{2}(a\!-\!a^{\ast}\!)
\!+\!\frac{i}{2}(b\!+\!b^{\ast}\!)\right]\overline{\psi}_{R}\boldsymbol{\gamma}^{\nu}\psi_{R}
\label{K}
\end{eqnarray}
which are field equations (\ref{Sciama}-\ref{Kibble}), respectively; then write torsion according to its most general decomposition given in the form
\begin{eqnarray}
&Q_{\rho\mu\nu}\!=\!\frac{1}{3}(g_{\rho\mu}Q_{\nu}\!-\!g_{\rho\nu}Q_{\mu})
\!+\!\frac{1}{6}W^{\beta}\varepsilon_{\beta\rho\mu\nu}\!+\!T_{\rho\mu\nu}
\label{decombination}
\end{eqnarray}
where $Q_{\nu}$ is of course the trace vector and $W^{\beta}$ the axial vector dual of the completely antisymmetric part, and while the remaining part is called $T_{\rho\mu\nu}$ such that it verifies contraction and permutation properties $T^{\rho}_{\phantom{\rho}\rho\nu}\!=\!0$ and $T_{\rho\mu\nu}\!+\!T_{\mu\nu\rho}\!+\!T_{\nu\rho\mu}\!=\!0$ with its independent contraction $T^{\mu\nu\rho}\varepsilon_{\alpha\beta\mu\nu}\!=\!-\frac{1}{2}T^{\rho\mu\nu} \varepsilon_{\alpha\beta\mu\nu}$ itself with its own contraction given by $T^{\rho\mu\nu}\varepsilon_{\alpha\rho\mu\nu}\!=\!0$ by construction, and which is such that $T^{\nu}_{\phantom{\nu}\alpha\sigma}\varepsilon^{\alpha\sigma\rho\mu}
\!+\!T^{\rho}_{\phantom{\rho}\alpha\sigma}\varepsilon^{\alpha\sigma\mu\nu}
\!+\!T^{\mu}_{\phantom{\mu}\alpha\sigma}\varepsilon^{\alpha\sigma\nu\rho}\!=\!0$ as a relationship that is valid in general, that is as a geometric identity. Now we may put all these results together in order to see that in fact the two independent field equations (\ref{Sciama}) and (\ref{Kibble}) are all that is present, as the field equation that corresponds to the irreducible non-axial part of torsion is given by the condition $T_{\rho\mu\nu}\!=\!0$ identically.

With these equations and definitions, we may now start by taking field equations (\ref{SK}) in which the source given by the current densities of the left-handed and right-handed semi-spinor fields are substituted with field equations (\ref{S}-\ref{K}); in this way, one obtains an expression that contains only torsion terms, which can be replaced in terms of the decomposition (\ref{decombination}): thus done, all the axial vector and the trace vector contributions disappear leaving only the irreducible non-axial terms in an expression given by
\begin{eqnarray}
\nonumber
&J(T_{\sigma\alpha}^{\phantom{\sigma\alpha}\rho}\varepsilon^{\sigma\alpha\mu\nu}
\!+\!\frac{1}{2}T^{\nu}_{\phantom{\nu}\sigma\alpha}\varepsilon^{\sigma\alpha\rho\mu}
\!-\!\frac{1}{2}T^{\mu}_{\phantom{\nu}\sigma\alpha}\varepsilon^{\sigma\alpha\rho\nu})-\\
\nonumber
&-I(T_{\alpha\sigma}^{\phantom{\alpha\sigma}\nu}\varepsilon^{\mu\rho\alpha\sigma}
-T_{\alpha\sigma}^{\phantom{\alpha\sigma}\mu}\varepsilon^{\nu\rho\alpha\sigma})
+FT^{\rho}_{\phantom{\rho}\alpha\sigma}\varepsilon^{\alpha\sigma\mu\nu}+\\
\nonumber
&+2YT^{\rho\mu\nu}\!+\!Z(T^{\nu\mu\rho}\!-\!T^{\mu\nu\rho})\!=\!0
\end{eqnarray}
or by considering the contraction and permutation properties
\begin{eqnarray}
&(2F-2J+I)T^{\rho}_{\phantom{\rho}\sigma\alpha}\varepsilon^{\sigma\alpha\mu\nu}
+2(2Y\!+\!Z)T^{\rho\mu\nu}\!=\!0
\label{auxiliary}
\end{eqnarray}
and multiplying by $\varepsilon_{\mu\nu\theta\eta}$ the equivalent form
\begin{eqnarray}
&-4(2F\!-\!2J\!+\!I)T^{\rho}_{\phantom{\rho}\theta\eta}
+2(2Y\!+\!Z)T^{\rho\mu\nu}\varepsilon_{\mu\nu\theta\eta}\!=\!0
\label{auxiliaryeq}
\end{eqnarray}
as it is easy to check. By considering now the equations (\ref{auxiliary}-\ref{auxiliaryeq}) we see that they can be simultaneously solved giving the constraint
\begin{eqnarray}
\nonumber
&[(2F\!-\!2J\!+\!I)^{2}\!+\!(2Y\!+\!Z)^{2}]T_{\rho\mu\nu}\!=\!0
\end{eqnarray}
and therefore the eventual
\begin{eqnarray}
\nonumber
&T_{\rho\mu\nu}\!=\!0
\end{eqnarray}
as we wanted to prove; this holds in general, unless we have the two conditions $2F\!-\!2J\!+\!I=0$ and $2Y\!+\!Z=0$ but in this case the tensor $T_{\rho\mu\nu}$ would disappear from the initial action, that is the non-axial irreducible part of torsion will not be dynamical and it will encode no physical degree of freedom. Thus, either the non-axial irreducible part of torsion $T_{\rho\mu\nu}$ is not present at all, or it is present but then the field equations set it to zero identically, as we wanted to prove in the most general terms, concluding the proof we intended to carry out in a detailed manner in this appendix.
\section{Proof of the Fierz identities}
\label{a2}
Along the paper we have also made use of the Fierz identities, and thus we give a general proof of those too: the very first thing we need to get is a way to express the product $\psi\overline{\psi}$ and since such a matrix belongs to the space of the complex $4\times4$ matrices, spanned by the $16$ linearly independent $\mathbb{I}$, $\boldsymbol{\pi}$, $\boldsymbol{\gamma}^{i}$, $\boldsymbol{\gamma}^{i}\boldsymbol{\pi}$, $\boldsymbol{\sigma}^{ij}$ matrices, we have that such a product can be expressed in terms of the linear combination given by $\psi\overline{\psi}\!=\!\phi\mathbb{I}\!+\!\omega\boldsymbol{\pi}
\!+\!A_{i}\boldsymbol{\gamma}^{i}\!+\!V_{i}\boldsymbol{\gamma}^{i}\boldsymbol{\pi}
\!+\!\frac{1}{2}S_{ij}\boldsymbol{\sigma}^{ij}$ in terms of $16$ complex functions we have to obtain: we observe that when this form is multiplied in turn by the $\mathbb{I}$, $\boldsymbol{\pi}$, $\boldsymbol{\gamma}^{i}$, $\boldsymbol{\gamma}^{i}\boldsymbol{\pi}$, $\boldsymbol{\sigma}^{ij}$ matrices and then taken in its trace, we get the five correspondent conditions that specify the functions $\phi$, $\omega$, $A_{i}$, $V_{i}$, $S_{ij}$, so that we get
\begin{eqnarray}
\nonumber
&\psi\overline{\psi}\!=\!\frac{1}{4}\overline{\psi}\psi\mathbb{I}
\!-\!\frac{1}{2}\overline{\psi}\boldsymbol{\sigma}_{ij}\psi\boldsymbol{\sigma}^{ij}
\!-\!\frac{1}{4}i\overline{\psi}\boldsymbol{\pi}\psi i\boldsymbol{\pi}+\\
&+\frac{1}{4}\overline{\psi}\boldsymbol{\gamma}_{i}\psi\boldsymbol{\gamma}^{i}
\!-\!\frac{1}{4}\overline{\psi}\boldsymbol{\gamma}_{i}\boldsymbol{\pi}\psi
\boldsymbol{\gamma}^{i}\boldsymbol{\pi}
\label{prove}
\end{eqnarray}
as it is easy to check; identity (\ref{prove}) can also be proven by choosing a representation and plugging the expression for the spinor and matrices and performing calculations, noticing that the result does not depend on the chosen representation. This is a general result.

Now expression (\ref{prove}) can be used to rearrange bi-linear spinor fields, as for instance in the expression for the scalar product between the vector and pseudo-vector currents of the spinor fields
\begin{eqnarray}
\nonumber
&\overline{\psi}\boldsymbol{\gamma}^{a}\psi\overline{\psi}
\boldsymbol{\gamma}_{a}\boldsymbol{\pi}\psi=\\
\nonumber
&=\overline{\psi}\boldsymbol{\gamma}^{a}(\frac{1}{4}\overline{\psi}\psi\mathbb{I}
\!-\!\frac{1}{2}\overline{\psi}\boldsymbol{\sigma}_{ij}\psi\boldsymbol{\sigma}^{ij}
\!-\!\frac{1}{4}i\overline{\psi}\boldsymbol{\pi}\psi i\boldsymbol{\pi}+\\
\nonumber
&+\frac{1}{4}\overline{\psi}\boldsymbol{\gamma}_{i}\psi\boldsymbol{\gamma}^{i}
\!-\!\frac{1}{4}\overline{\psi}\boldsymbol{\gamma}_{i}\boldsymbol{\pi}\psi
\boldsymbol{\gamma}^{i}\boldsymbol{\pi})\boldsymbol{\gamma}_{a}\boldsymbol{\pi}\psi=\\
\nonumber
&=\overline{\psi}\psi\overline{\psi}\boldsymbol{\pi}\psi
\!+\!i\overline{\psi}\boldsymbol{\pi}\psi i\overline{\psi}\psi-\\
\nonumber
&-\frac{1}{2}\overline{\psi}\boldsymbol{\gamma}_{i}\psi
\overline{\psi}\boldsymbol{\gamma}^{i}\boldsymbol{\pi}\psi
\!-\!\frac{1}{2}\overline{\psi}\boldsymbol{\gamma}_{i}\boldsymbol{\pi}\psi
\overline{\psi}\boldsymbol{\gamma}^{i}\psi\equiv\\
\nonumber
&\equiv-\overline{\psi}\boldsymbol{\gamma}_{i}\psi
\overline{\psi}\boldsymbol{\gamma}^{i}\boldsymbol{\pi}\psi
\end{eqnarray}
since $\boldsymbol{\gamma}^{a}\boldsymbol{\gamma}_{a}\!=\!4\mathbb{I}$ and $\boldsymbol{\gamma}^{a}\boldsymbol{\gamma}^{i}\boldsymbol{\gamma}_{a}
\!=\!-2\boldsymbol{\gamma}^{i}$ and $\boldsymbol{\gamma}^{a}\boldsymbol{\sigma}^{ij}\boldsymbol{\gamma}_{a}\!=\!0$ as the well known properties of these matrices in $(1\!+\!3)$-dimensional spacetimes and for the $4\times4$ complex representation space, and therefore
\begin{eqnarray}
\nonumber
&\overline{\psi}\boldsymbol{\gamma}^{a}\psi\overline{\psi}
\boldsymbol{\gamma}_{a}\boldsymbol{\pi}\psi\!=\!0
\end{eqnarray}
as it may also be proven with direct calculations; in a similar manner, one may also demonstrate that the squares of these currents are related by the following constraining condition
\begin{eqnarray}
\nonumber
&\overline{\psi}\boldsymbol{\gamma}^{a}\psi\overline{\psi}\boldsymbol{\gamma}_{a}\psi
\!=\!-\overline{\psi}\boldsymbol{\gamma}_{a}\boldsymbol{\pi}\psi\overline{\psi}
\boldsymbol{\gamma}^{a}\boldsymbol{\pi}\psi
\end{eqnarray}
as computations would have shown explicitly. This method is general, and it can be used to get more relationships systematically.

A list of some of the most important of these identities is as: 
\begin{eqnarray}
&\overline{\psi}\boldsymbol{\gamma}_{a}\psi\overline{\psi}\boldsymbol{\gamma}^{a}\psi
\!=\!|\overline{\psi}\psi|^{2}\!+\!|i\overline{\psi}\boldsymbol{\pi}\psi|^{2}
\!\!=\!-\overline{\psi}\boldsymbol{\gamma}_{a}\boldsymbol{\pi}\psi
\overline{\psi}\boldsymbol{\gamma}^{a}\boldsymbol{\pi}\psi
\label{id1}\\
&\overline{\psi}\boldsymbol{\gamma}_{a}\boldsymbol{\pi}\psi
\overline{\psi}\boldsymbol{\gamma}^{a}\psi=0
\label{id2}
\end{eqnarray}
together with the complementary
\begin{eqnarray}
&2i\overline{\psi}\boldsymbol{\sigma}_{ab}\psi i\overline{\psi}\boldsymbol{\sigma}^{ab}\psi
\!=\!|\overline{\psi}\psi|^{2}\!-\!|i\overline{\psi}\boldsymbol{\pi}\psi|^{2}
\!\!=\!2i\overline{\psi}\boldsymbol{\sigma}_{ab}\boldsymbol{\pi}\psi
i\overline{\psi}\boldsymbol{\sigma}^{ab}\boldsymbol{\pi}\psi\\
&i\overline{\psi}\boldsymbol{\sigma}_{ab}\boldsymbol{\pi}\psi
\overline{\psi}\boldsymbol{\sigma}^{ab}\psi
=-i\overline{\psi}\boldsymbol{\pi}\psi\overline{\psi}\psi
\end{eqnarray}
and also the following
\begin{eqnarray}
&2i\overline{\psi}\boldsymbol{\sigma}_{ik}\psi
\overline{\psi}\boldsymbol{\gamma}^{i}\psi
=i\overline{\psi}\boldsymbol{\pi}\psi\overline{\psi}\boldsymbol{\gamma}_{k}\boldsymbol{\pi}\psi\\
&2\overline{\psi}\boldsymbol{\pi}\boldsymbol{\sigma}_{ik}\psi
\overline{\psi}\boldsymbol{\gamma}^{i}\psi
=\overline{\psi}\psi\overline{\psi}\boldsymbol{\gamma}_{k}\boldsymbol{\pi}\psi\\
&2i\overline{\psi}\boldsymbol{\sigma}_{ik}\psi
\overline{\psi}\boldsymbol{\gamma}^{i}\boldsymbol{\pi}\psi
=i\overline{\psi}\boldsymbol{\pi}\psi\overline{\psi}\boldsymbol{\gamma}_{k}\psi\\
&2\overline{\psi}\boldsymbol{\pi}\boldsymbol{\sigma}_{ik}\psi
\overline{\psi}\boldsymbol{\gamma}^{i}\boldsymbol{\pi}\psi
=\overline{\psi}\psi\overline{\psi}\boldsymbol{\gamma}_{k}\psi
\end{eqnarray}
together with
\begin{eqnarray}
&\overline{\psi}\psi i\overline{\psi}\boldsymbol{\sigma}_{ab}\psi\!-\!
i\overline{\psi}\boldsymbol{\pi}\psi\overline{\psi}\boldsymbol{\sigma}_{ab}\boldsymbol{\pi}\psi
\!=\!\frac{1}{2}\overline{\psi}\boldsymbol{\gamma}^{j}\psi
\overline{\psi}\boldsymbol{\gamma}^{k}\boldsymbol{\pi}\psi\varepsilon_{jkab}
\end{eqnarray}
again as they can be checked with (\ref{prove}) or by calculation explicitly.

When decomposed for the left-handed and right-handed chiral projections, some of these identities reduce to the simpler
\begin{eqnarray}
&\!\!\!\!2\overline{\psi}_{L}\boldsymbol{\gamma}_{a}\psi_{L}
\overline{\psi}_{R}\boldsymbol{\gamma}^{a}\psi_{R}\!\!=\!
|\overline{\psi}_{L}\psi_{R}\!+\!\overline{\psi}_{R}\psi_{L}|^{2}
\!\!-\!|\overline{\psi}_{L}\psi_{R}\!-\!\overline{\psi}_{R}\psi_{L}|^{2}\\
&\overline{\psi}_{L}\boldsymbol{\gamma}_{a}\psi_{L}
\overline{\psi}_{L}\boldsymbol{\gamma}^{a}\psi_{L}\!=\!\overline{\psi}_{R}\boldsymbol{\gamma}_{a}\psi_{R}
\overline{\psi}_{R}\boldsymbol{\gamma}^{a}\psi_{R}\!=\!0
\end{eqnarray}
as it can be checked by decomposing (\ref{id1}-\ref{id2}) in left-handed and right-handed chiral semi-spinorial projections of the spinor, and these are precisely the types of Fierz identities that we have been employing to rearrange the components of the left-handed and right-handed semi-spinorial matter field throughout the paper.

As it is clear, many more of such identities actually exist, all of them obtained by either writing a possible square of bilinear spinors and then applying (\ref{prove}) or directly proving them with an explicit decomposition, but for purpose we have in the present paper the identities we have listed are more than enough.

\end{document}